\documentclass[preprint,showpacs,preprintnumbers,amsmath,amssymb]{revtex4}

\usepackage{graphicx}
\usepackage{dcolumn}
\usepackage{bm}

\begin{document}
\title{About the two spin-channel model for ferromagnetic excitations and spin-dependent heat transfer equations}
 \author{J.-E. Wegrowe} \email{jean-eric.wegrowe@polytechnique.fr}
 \affiliation{Ecole Polytechnique, LSI, CNRS and CEA/DSM/IRAMIS, Palaiseau F-91128, France}
 
\date{\today}

\begin{abstract}
The two spin-channel model is generalized to the case of transport of ferromagnetic excitations in electric conductors and insulators. The two channels are defined by reducing the ferromagnetic degrees of freedom to a bivaluated variable, i.e. to an effective spin one-half. The reduction is performed after defining the local magnetic configuration space by a sphere $\Sigma_x$, and  integrating the relevant physical quantities over the two hemispheres $\Sigma_x^{\uparrow}$ and  $\Sigma_x^{\downarrow}$. The configuration space is then extended to the $x$ direction for non-uniform magnetization excitations. The transport equations for both magnetic moments and magnetic energy are deduced, including the relaxation from one channel to the other. The heat transport equations for ferromagnets is deduced.
\end{abstract}

\pacs{ 75.47.-m, 72.25.-b, 85.80.Lp \hfill}

\maketitle

In the context of spintronics, the effect of spin-injection and spin-accumulation is easy to describe on the basis of the two spin-channel model for electric carriers, if the magnetization is locally defined by the microscopic spin one-half $s = \pm \hbar/2$  \cite{Fert,Aronov,Johnson,Wyder,Valet}. In that case, the electric carriers define without ambiguity the two channels at any points $x$ of the material: one channel for the up spin $\hbar/2$ and the other channel for the down spin $- \hbar/2$, with a fixed quantification axis. However, in the case of ferromagnetic interactions between electric carriers (e.g. for spin-waves excitations), this definition is {\it a-priori} not valid since the magnetic carriers are not longer defined locally by a spin one -half. Beyond, in the case of electric insulators, the definition of two spin-channel seems to be problematic because it cannot be based on an assembly of delocalized quasi-particles.

\begin{figure}
   \begin{center}
   \begin{tabular}{c}
   \includegraphics[height=7cm]{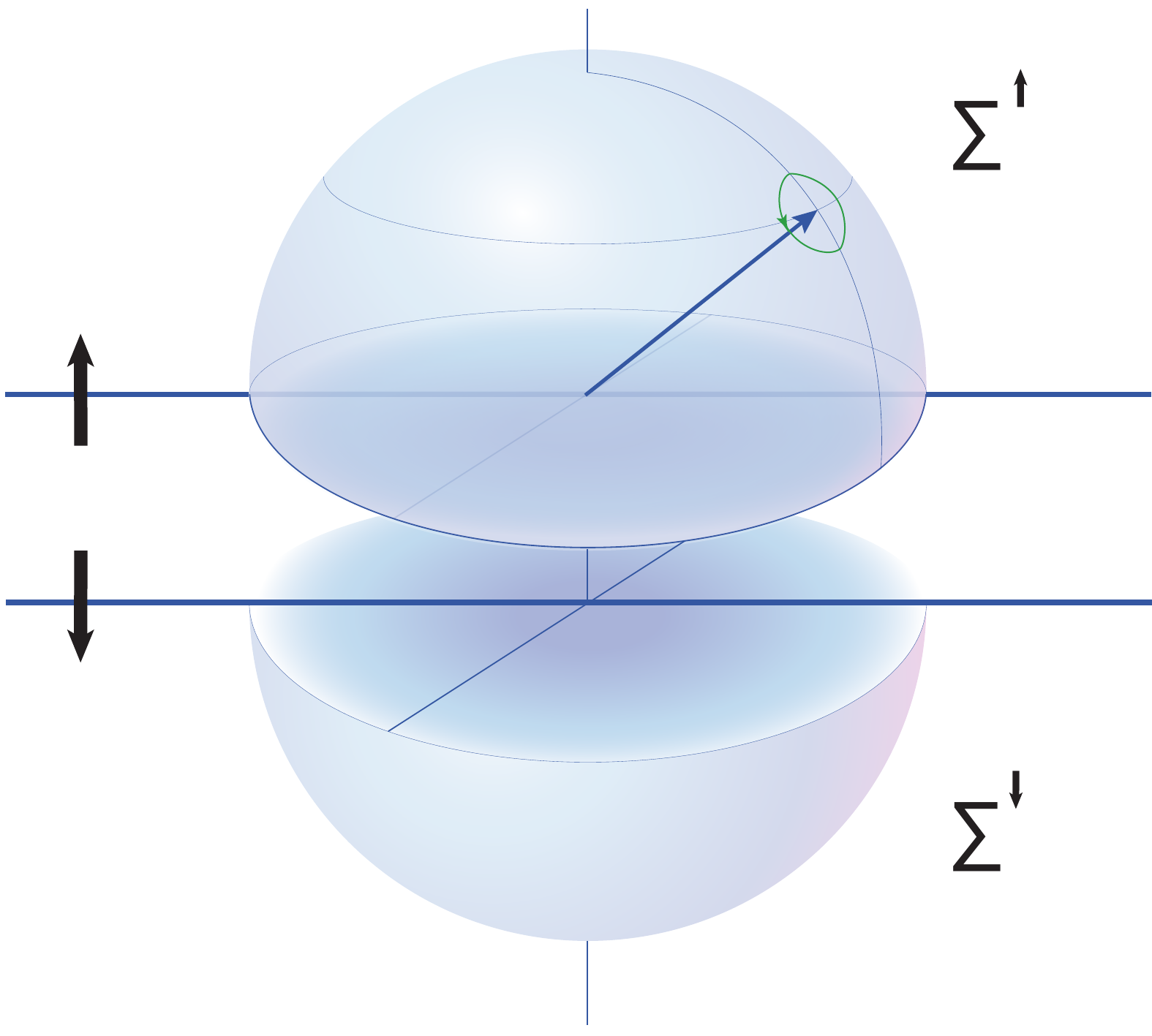}
   \end{tabular}
   \end{center}
   \caption[Hemisphere] 
{ \label{fig:Hemisphere} Illustration of the $\Sigma_x$ configuration space of ferromagnetic degrees of freedom at point $x$. The separation in two hemispheres  $\Sigma_x = \Sigma^{\uparrow}_x \cup \Sigma^{\downarrow}_x$ allows the two effective spin-channels to be defined. The arrow in the center represents a mean value of the magnetization in the configuration space.}
   \end{figure}

We show in this paper that the two spin-channel model can nevertheless be generalized to any kind of macroscopic ferromagnetic excitations (quantal or classical). This generalization is based on a reduction method that allows the continuous magnetic degrees of freedom to be reduced to a bivaluated variable at each point $x$ of the usual space, i.e. to a local effective ``spin" one-half. The reduction is performed after defining the local magnetic configuration space over the sphere $\Sigma_x$ (section I below), and integrating the relevant physical quantities over each hemisphere  (see Fig. 1) for each point $x$ of the usual space (section II). The configuration space is then extended to the $x$ direction for non-uniform magnetization excitations, and the transport equations are deduced (section III).

This generalization of the two spin-channel model shows that the effects generated by spin-polarized heat currents are similar to that generated by electric spin-polarized currents. Typically, the spin-accumulation effect occurring in spintronics devices can be generalized to {\it thermal spin-accumulation} \cite{CondMat}. The similarity of the transport equations of heat - together with the usual Seebeck and Nernst effects - could explain the so-called spin-Seebeck and spin-Nernst effects observed recently on various materials \cite{Uchida, Jaworsky,Uchida2,Sharoni,Rezende,Huang,Chumak}. 

\subsection{Definition of the localized ferromagnetic configuration space $\Sigma_x$}

In this section, we focus on the localized uniform ferromagnetic
moment $\vec M(x) = M_{s} \vec e_{r}(x)$ defined with radial unit vector
$\vec e_{r}(x)$ and magnetization at saturation $M_{s}$ at point $x$. The approach used is the mesoscopic non-equilibrium thermodynamic theory (MNET) \cite{Mazur,Regera} applied to rotational brownian motion \cite{Brown,Coffey}. 

 In order to treat statistically the ferromagnetic degrees of freedom, a statistical ensemble of a large number of ferromagnetic moments is defined on the configuration space $\Sigma_x$ (Fig. 1).  Each magnetic moment is described by its position $\{ \theta, \varphi \}$ on the sphere $\Sigma_x$ of radius $M_{s}$. The angle $\theta$ is associated to the radial unit vector $\vec e_{\theta}$ and the angle $  \varphi$ is associated to the azimuth unit vector $\vec e_{\varphi}$.

 The statistical distribution of the magnetic moments on the sphere $\Sigma_x$ is then defined by the density $\rho^{F}(\theta,\varphi;x)$ per units of solid angle $d\Omega = sin \theta \, d\theta \, d \varphi$ \cite{PRB08} and per unit length $\delta x$. The function $\rho^{F}(\theta,\varphi)$ is a solution of the rotational Fokker-Planck equation derived at the end of this subsection. At equilibrium, the density is given by the Boltzmann distribution defined by the ferromagnetic potential $V^{F}(\theta, \varphi; x)$, and such that the effective magnetic field vanishes $\vec H_{eff} = - \vec \nabla V^F = 0$ ($\vec H_{eff}$ contains all deterministic contributions: external magnetic field, dipolar field, anisotropy field, exchange field, etc). For out-of-equilibrum states, fluctuations and diffusion plays a fundamental role. The distribution is given by the ferromagnetic chemical potential that takes the form $\mu^{F} = kT \, ln(\rho^{F}) + V^{F}$  \cite{Mazur,Raikher,PRB08}.  The generalized force $\vec H_{eff} + \vec h \equiv -\vec \nabla_{\Sigma} \mu^{F}$ produces a current of
ferromagnetic moments $\vec J^{F}(\theta, \varphi;x) = \rho^F d \vec{u}_{r}/dt$, that is flowing on the surface of the sphere. This generalized force contains 
a deterministic part $\vec H_{eff} = - \vec \nabla V^{F}$ and a diffusive part $\vec h = ( kT/\rho^F ) \vec \nabla \rho^F$  \cite{Brown,Raikher}.

 If the ferromagnetic system is closed, the magnetic moments are conserved on $\Sigma_x$ so that:
\begin{equation}
\frac{d \rho^F}{dt} = - div_{\Sigma} \vec J^F.
\label{ContinuityEq0}
\end{equation}

On the other hand, the ferromagnetic energy $u^F$ is also described by a the continuity equation. However, the system under interest is {\it not adiabatic} and the conservation equation for the ferromagnetic energy takes the form: 
\begin{equation}
\frac{d u^F}{dt} = - div_{\Sigma} \vec J^u + r
\label{ContinuityU}
\end{equation}
where $r$ describes the dissipation in the environment and $\vec J^u$ is the flux of ferromagnetic energy in the configuration space $\Sigma_x$.  If the temperature $T(x)$ is uniform over the configuration space $\Sigma_x$ (i.e. at each point $x$), the flux of heat is zero, and the power dissipated by the ferromagnetic system is reduced to the effect of the flux of magnetic moments:  the ferromagnetic power dissipated or stored at each point $x$ of the ferromagnet reads
\begin{equation}
\mathcal P^F= - \int_\Sigma \vec J^{F} . \vec \nabla_{\Sigma} \mu^{F} d\Omega
\label{Power}
\end{equation}
The application of the second law of thermodynamics
allows the transport equation to be deduced
by writing the relation that links the generalized flux to 
the generalized
force.  Both quantities, 
flux and
forces, are then related by the Onsager matrix of the transport
coefficients $\bar{\mathcal L}$ \cite{PRB08,SSC10,Entropy11}:
\begin{equation}
    \vec J^{F} = - \bar{\mathcal L} \, \vec \nabla_{\Sigma} 
    \mu^{F}
   \label{Jferro}
\end{equation}

where the flow $\vec J^F$ is a two component
vector defined with the unit vectors $\{ \vec e_{\varphi}, \vec
e_{\theta} \}$ of $\Sigma_x$.  Accordingly, the Onsager matrix is a 2x2 matrix
defined by four transport coefficients $\{L_{\theta \theta},
L_{\theta \varphi}, L_{\varphi \theta}, L_{\varphi \varphi} \}$. 
The Onsager reciprocity relations impose that $L_{\theta
\varphi} = - L_{\varphi \theta}$. Furthermore, assuming that the dissipation is 
isotropic, we have $L_{\theta \theta} = L_{\varphi \varphi}$.  Introducing $
\alpha$ as the ratio of the off-diagonal to the diagonal coefficients;
$ \alpha = L_{\theta \varphi} / L_{\theta \theta} $, the ferromagnetic kinetic equation is defined by two ferromagnetic transport coefficients $ L_{F} = L_{\theta \varphi}/\rho^{F} $ and $ \alpha$:
\begin{equation}
\bar{\mathcal L} =
\rho^{F} L_{F} \,  \left( \begin{array}{cc}
 \alpha &  1 \\
                        - 1 & \alpha \\
\end{array} \right)
\label{MatrixL0}
\end{equation}

Rewriting Eq. (\ref{Jferro}) in the
reference frame $\{ \vec e_{r}, \vec e_{\theta}, \vec e_{\varphi} \}$,
and recalling that the current is the density multiplied by the
velocity we obtain the well known LL equation:

\begin{equation}
\frac{d\vec e_{r}}{dt}  =
- L_{F} \, \left \{ 
 \vec e_{r} \times \left ( \vec H_{eff} +  \frac{kT}{\rho} \vec \nabla \rho  \right ) + \alpha \vec e_{r} \times \left ( 
\vec e_{r} \times
 \left ( \vec H_{eff} +  \frac{kT}{\rho} \vec \nabla \rho  \right ) \right ) \right \} 
. 
 \label{LL}
\end{equation}
The equivalence between the LL equation and the
phenomenological Gilbert equation \cite{AmJPh} 
gives the relation between the coefficients $\alpha$ and $L_{F}$ on the one hand, and the Gilbert coefficients 
$\eta$ and the gyromagnetic ratio $\gamma$ on the other hand: 
\begin{equation}
\begin{array}{c}
\alpha = \eta \gamma M_{s} \\
         L_{F}
		 =  \frac{\gamma}{M_{s} \left (1+ \alpha^{2} \right )}
\label{Ltheta}
\end{array}
\end{equation}

Inserting Eq. (\ref{MatrixL0}) and Eq. Eq. (\ref{Jferro}) into the continuity equation Eq. (\ref{ContinuityEq0}) leads to the Fokker-Planck equation in the configuration space $\Sigma_x$ :
\begin{equation}
\frac{d \rho^F}{dt} =  \vec \nabla_{\Sigma} .  \left \{ 
\rho^F  \vec e_{r} \times \left (L_{F} \vec H_{eff} +  D \vec \nabla_{\Sigma} \rho^F  \right ) + \alpha \rho^F \, \vec e_{r} \times \left ( 
\vec e_{r} \times
 \left (L_{F} \vec H_{eff} +  D \vec \nabla_{\Sigma} \rho^F  \right )
 \right ) \right \} . 
\label{LLG}
\end{equation}
where $D = L_F kT/\rho^F $ is the diffusion coefficient.
This result is well known \cite{Brown,Coffey}. The goal is to generalize the description to non-uniform ferromagnets (i.e. performing the extension of the configuration space to the neighbors $\Sigma_{x \pm \delta x}$, in order to describe transfer of magnetic moments (beyond Eq. (\ref{ContinuityEq0})) and transfer of energy. The objective of the next section is to simplify the problem by defining fist the two channel model for magnetic excitations.  


\subsection{The two spin-channel model: reduction of the ferromagnetic degrees of freedom to a bivaluted variable.}
The concept of ``spin-channels" is inspired from spin-dependent transport studies (or ``spintronics"), for which the electronic spin is a bivaluated degree of freedom  with $s = \pm \hbar/2$ corresponding to the two states $\uparrow$ and $\downarrow$ for fixed quantization axis. 

We will show in the following the a ferromagnetic bivaluated variable can be defined by reducing the continuous degrees of freedom of the magnetization to a bivaluated variable, i.e. to an {\it  effective spin}. 
This reduction is performed by the integration of the relevant quantities over the two hemispheres $\Sigma^{\uparrow}_x$ and  $\Sigma^{\downarrow}_x$, such that $\Sigma_x = \Sigma^{\uparrow}_x \cup \Sigma^{\downarrow}_x$ (as shown in Fig. 1). We can then define the chemical potentials of the two hemispheres by:
\begin{equation}
\mu^F_{\updownarrow}(x) = \int_{\Sigma_x^{\updownarrow}} \mu(\theta, \varphi; x) d \Omega,
\end{equation}

 The total chemical potential is given by the sum over the two hemisphere $\mu^F = \mu^F_{\uparrow} + \mu^F_{\downarrow}$, and the difference defines a ``{\it pumping force}" \cite{Entropy11} $\Delta \mu = \mu^F_{\uparrow} - \mu^F_{\downarrow}$. 
The number of magnetic moments $n^F_{\updownarrow}$ per unit length $\delta x$ for each hemisphere reads:
\begin{equation}
n^F_{\updownarrow}(x) = \int_{\Sigma_x^{\updownarrow}} \rho^F(\varphi,\theta;x) d \Omega,
\label{Number}
\end{equation}
 so that the density of magnetic moments per unit length is $n^F_0 = n^F_{\uparrow} + n^F_{\downarrow}$ and the difference between the hemispheres is $\Delta n^F = n^F_{\uparrow} - n^F_{\downarrow}$.

Let us define the scalar $\mathcal J^F_{\updownarrow}$ by the expression:
\begin{equation}
\mathcal J^F_{\updownarrow}(x) = \int_{\Sigma_x^{\updownarrow}} div_{\Sigma}  \vec J^F  \, d \Omega, 
\label{DefJ}
\end{equation}

and the difference 

\begin{equation}
\Delta \mathcal J^F(x) =  \int_{\Sigma_x^{\uparrow}} div_{\Sigma}  \vec J^F  \, d \Omega -\int_{\Sigma_x^{\downarrow}} div_{\Sigma}  \vec J^F  \, d \Omega.
\label{dotpsi}
\end{equation}

From Eq. (\ref{ContinuityEq0}), Eq. (\ref{Number})  and Eq. (\ref{DefJ}) we have:
\begin{equation}
\begin{array}{c}
\frac{d n_{\uparrow}^F}{dt} =  \mathcal J^F_{\uparrow} = \frac{ \Delta \mathcal J^F}{2}\\
\frac{d n_{\downarrow}^F}{dt} = \mathcal J^F_{\downarrow} = - \frac{\Delta \mathcal J^F}{2}
\end{array}
\label{ContinuityEq1}
\end{equation} 

where the last equality in the left hand side is obtained by noting that the total numbers of magnetic moments on the sphere $n_0 = n_{\uparrow} + n_{\downarrow}$ is constant, so that $d n_0^F/dt =  \mathcal J^F_{\uparrow} + \mathcal J^F_{\downarrow}= 0$. 

Note that $\Delta \mathcal J^F$ is the flux of the magnetic moments flowing from the hemisphere $\Sigma^{\uparrow}_x$ to the other hemisphere $\Sigma^{\downarrow}_x$, and the ``{\it pumping force}" $\Delta \mu$ is thermodynamically conjugate to the ``flux" $\frac{\Delta \mathcal J^F}{2}$ in the sense that the product $\mathcal P^{\uparrow \, \downarrow} =  \Delta \mathcal J^F \Delta \mu /2$ is the ferromagnetic power exchanged between the two hemispheres.

In the same way as above, the ferromagnetic energy $u^F(\theta, \varphi;x)$ is integrated over the two hemispheres. The reduced variable reads:

\begin{equation}
\mathcal U^F_{\updownarrow}(x) = \int_{\Sigma_x^{\updownarrow}} u^F(\varphi,\theta;x) d \Omega,
\label{NumberU}
\end{equation}
and from Eq. (\ref{ContinuityU}) we have:

\begin{equation}
\begin{array}{c}
\frac{d \mathcal U^F_{\uparrow}}{dt} =  \mathcal J^\mathcal U_{\uparrow} + \mathcal R/2 \\
\frac{d \mathcal U^F_{\downarrow}}{dt} = \mathcal J^\mathcal U_{\downarrow} + \mathcal R/2 
\end{array}
\label{ContinuityEq2}
\end{equation} 

where $ \mathcal R/2 = \int_{\Sigma_{\uparrow}} r d\Omega = \int_{\Sigma_{\downarrow}} r d\Omega$, and the energy currents $\mathcal J^ \mathcal U_{\updownarrow}$ are defined on each point $x$ by the relation:
\begin{equation}
\mathcal J^\mathcal U_{\updownarrow}(x) = \int_{\Sigma_x^{\updownarrow}} div_{\Sigma}  \vec {J^{\mathcal U}}  \, d \Omega. 
\label{DefJu}
\end{equation}

If the system is adiabatic total energy $\mathcal U_0 = \mathcal U_{\uparrow} + \mathcal U_{\downarrow}$ on the sphere $\Sigma_x$ would constant and we would have, as for the current of magnetic moments:
\begin{equation}
\begin{array}{c}
\mathcal J^\mathcal U_{\uparrow} = \frac{ \Delta \mathcal J^\mathcal U}{2}\\
\mathcal J^\mathcal U_{\downarrow} = - \frac{\Delta \mathcal J^\mathcal U}{2},
\end{array}
\end{equation}
where $\Delta \mathcal J^\mathcal U= \mathcal J^\mathcal U_{\uparrow} - \mathcal J^\mathcal U_{\downarrow}$.
In conclusion, even for an isolated and uniform ferromagnetic particle (i.e. adiabatic ferromagnetic system), a current of energy is flowing from one hemisphere to the other at any point $x$. In other terms, an {\it effective ``spin-flip relaxation"} has been defined for ferromagnetic excitations. 

\subsection{Transport equations for heat current along the $x$ direction.}

A ferromagnetic wire can be modeled  by a succession of segments of thickness $\delta x$ and section unity  \cite{PRB00}. At each position $x$, the magnetization of volume $\pm \delta x$ is described in the configuration space $\Sigma_x$ with the corresponding chemical potential. The configuration space of the total ferromagnetic wire of length $l$ is then the continuous limit ($\delta x \rightarrow 0$,  $N \rightarrow \infty$ and $N \delta x = l$) of a chain of $N$ ferromagnetic configuration spaces $...\cup \Sigma_{x-\delta x} \cup \Sigma_x \cup \Sigma_{x+\delta x} \cup ...$  (see Fig. 2).

Let us first consider {\it an open system} that is able to exchange magnetic moments with its environment \cite{PRB00}. In this case, a mechanism of transport of magnetic moments would take place in the $x$ direction, and the conservation equations for magnetic moments through the space $x$ (Eq. (\ref{ContinuityEq1})) would take the same form as in the case of spin-dependent transport in the two channel model for electric conductors:
\begin{equation}
\begin{array}{c}
\frac{dn_{\uparrow}}{dt} =  - \frac{\partial \mathcal J^F_{\uparrow}}{\partial x}  - \dot \psi^F \\
\frac{dn_{\downarrow}}{dt} =  - \frac{\partial \mathcal J^F_{\downarrow}}{\partial x} + \dot \psi^F
\end{array}
\label{MagTransport}
\end{equation}
where $\dot \psi^F \equiv \Delta \mathcal J^F/ (2 \delta x)$ is the effective spin-flip relaxation, in the sense that the flux $\dot \psi^F$ is the velocity of the transformation of ``spin up" $\uparrow$ into a ``spin down" $\downarrow$ under the action of the chemical affinity $\Delta \mu^F$ \cite{Wyder, Valet,PRB00,PRB08,SSC10,Entropy11,Inertia}.

\begin{figure}
   \begin{center}
   \begin{tabular}{c}
   \includegraphics[height=4cm]{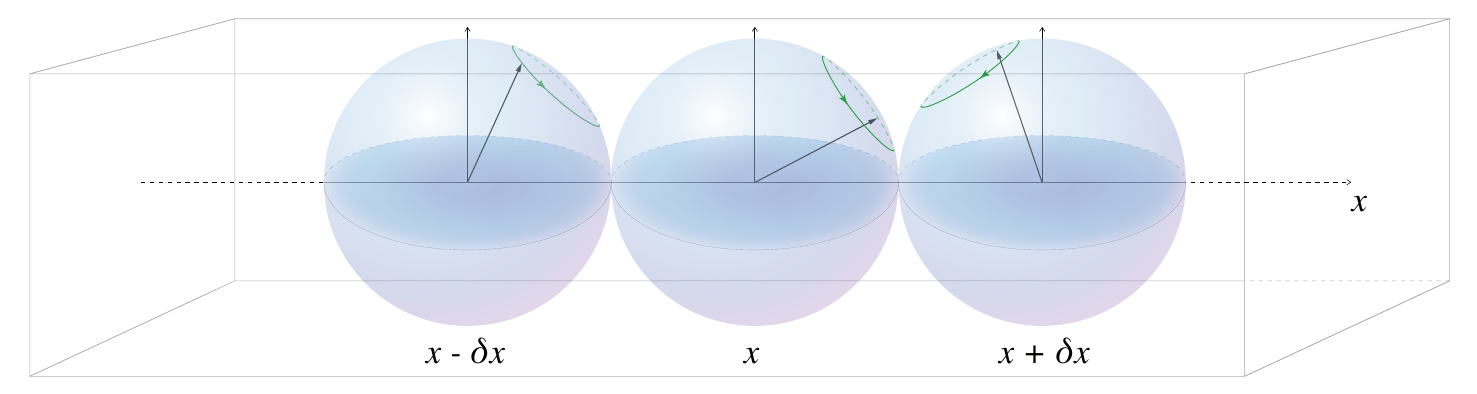}
   \end{tabular}
   \end{center}
   \caption[Trisphere] 
{ \label{fig:Trisphere} Illustration of the configuration space $\Sigma$ for a non-uniform
ferromagnet of length $l = N \delta x$ and section unity, defined as a chain of $N$ spheres: $\Sigma = ...\cup \Sigma_{x-\delta x} \cup \Sigma_x \cup \Sigma_{x+\delta x} \cup ...$  at the continuous limite ($N \rightarrow \infty, \delta x \rightarrow 0$).}
  \end{figure}

However, in a ferromagnet, magnetic moments are localized so that a spin-injection mechanism carried by the magnetic excitations is not expected. The magnetic moments cannot be transmitted from the configuration space $\Sigma^{\updownarrow}_x$ to its neighbors $\Sigma^{\updownarrow}_{x\pm 1}$ and {\it the local configuration space is closed} from the point of view of the ferromagnetic degrees of freedom. 

In contrast, {\it since the system under interest is not adiabatic} energy is transmitted from the configuration space $\Sigma^{\updownarrow}_x$ to its neighbors $\Sigma^{\updownarrow}_{x\pm 1}$. Typically, this situation corresponds to ferromagnetic resonance or heating performed at an extremity of the ferromagnetic device \cite{Uchida, Jaworsky,Uchida2,Sharoni,Rezende,Huang,Chumak}. Space-dependent and spin-dependent heat currents  $\mathcal J^\mathcal U_{\updownarrow}(x)$ are then produced and flow throughout the sample and through the interfaces. Furthermor, as shown in Eq. (\ref{ContinuityEq2}), the energy flux $\Delta \mathcal J^{\mathcal U}$ is produced inside the configuration space $\Sigma_x$, flowing from the subspace $\Sigma^{\uparrow}_x$ to the subspace $\Sigma^{\downarrow}_x$. Generalizing Eq. (\ref{ContinuityEq2}) with the contribution of the neighbors $\Sigma_{x \pm 1}$, we have:
\begin{equation}
\begin{array}{c}
\frac{d \mathcal U_{\uparrow} }{dt} = - \frac{ \partial \mathcal J^\mathcal U_{\uparrow}}{\partial x} - \dot  \psi^{th}  +  \mathcal R/2 \\
\frac{d\mathcal U_{\downarrow}}{dt} = - \frac{ \partial \mathcal J^\mathcal U_{\downarrow}}{\partial x} + \dot \psi^{th} + \mathcal R/2,
\end{array}
\end{equation}
where $\dot \psi^{th} \equiv \frac{\Delta \mathcal J^\mathcal U}{2\delta x}$ is due to the relaxation between the two channels. 
The heat current generated by the energy carriers is given by the relation:
\begin{equation}
 \mathcal J^q_{\updownarrow} = \mathcal J^\mathcal U_{\updownarrow} - \mu_{\updownarrow}  \mathcal J^F_{\updownarrow}
\label{HeatCurrent}
\end{equation}
We are here interested in the case $J^F_{\updownarrow} = 0$. The diffusion equations for each channel can then be deduced \cite{CondMat}:
\begin{equation}
\mathcal J^q_{\updownarrow} = - \mathcal L_{\updownarrow} \frac{\partial \mu^F_{\updownarrow}}{\partial x} + \lambda_{\updownarrow} \frac{\partial T}{\partial x}
\label{HeatOnsager}
\end{equation}
\subsection{Conclusion}
A non-equilibrium thermodynamic approach has been used in order to establish a two channel model applied to macroscopic ferromagnetic excitations. The reduction of the ferromagnetic degrees of freedom from two continuous coordinates $\{ \theta, \varphi \}$ to a bivaluated scalar variable $\updownarrow$ (and a quantification axis) is explicitly performed, and applied to the relevant physical observables, i.e. the density of magnetic moments, the energy density, and the corresponding currents. The complexity of the transport properties of magnetic moments and magnetic energy (including diffusion) can be reduced to the kinetic equations of two states $\uparrow$ and $\downarrow$ at each position $x$ in space. This method allows the spintronics concept of  {\it spin-injection} to be generalized to any kind of magnetic excitations. In particular, it is shown that {\it thermal spin-injection} can be performed without magnetic carriers, but with spin-dependent heat currents only. The validity of the usual technics of spintronics has been extended to the case of ferromagnetic excitations (quantal or classical) occurring in electric conductors or electric insulators.

\end{document}